\documentclass[times]{asjcauth}
\usepackage{graphicx }
\usepackage{amssymb}
\usepackage{mathrsfs}
\usepackage{enumerate}
\usepackage{multirow}
\usepackage{booktabs}
\usepackage{moreverb}
\usepackage[colorlinks,
            bookmarksopen,
            bookmarksnumbered,
            linkcolor=blue,
            anchorcolor=blue,
            citecolor=blue]{hyperref}
\usepackage[numbers,sort&compress]{natbib}
\usepackage[misc]{ifsym}
\newtheorem{theorem}{Theorem}

\newtheorem{remark}{Remark}                             %

\begin{document}

\runninghead{Y. H. Wei: Modelling and simulation of nabla fractional dynamic systems $\cdots$}


\title{Modelling and simulation of nabla fractional dynamic systems\\ with nonzero initial conditions}

\author{Yiheng Wei, Jiachang Wang, Peter W Tse, Yong Wang$^*$}
\address{Y. Wei, J. Wang and Y. Wang (\Letter) are with the Department of Automation, University of Science and Technology of China, Hefei, 230026, China. Email: yongwang@ustc.edu.cn. P. W. Tse is with the Department of Systems Engineering and Engineering Management, City University of Hong Kong, Hong Kong, 999077, China.}


\acks{The work described in this paper was fully supported by the National Natural Science Foundation of China (61601431, 61573332), the Anhui Provincial Natural Science Foundation (1708085QF141) and the fund of China Scholarship Council (201806345002). This draft was updated after it was accepted by Asian Journal of Control.}

\begin{abstract}
The paper focuses on the numerical approximation of nabla fractional order systems with the conditions of nonzero initial instant and nonzero initial state. First, the inverse nabla Laplace transform is developed and the equivalent infinite dimensional frequency distributed models of discrete fractional order system are introduced. Then, resorting the nabla Laplace transform, the rationality of the finite dimensional frequency distributed model approaching the infinite one is illuminated. Based on this, an original algorithm to estimate the parameters of the approximate model is proposed with the help of vector fitting method. Additionally, the applicable object is extended from a sum operator to a general system. Three numerical examples are performed to illustrate the applicability and flexibility of the introduced methodology.
\end{abstract}


\maketitle

\section{Introduction}\label{Section 1}
Fractional calculus is a natural generalization of classical integer order calculus, whose inception can be traced back to 300 years ago. It is well known that fractional calculus has been widely applied to system modelling \cite{Oustaloup:2013Auto,Song:2017AJC}, stability analysis \cite{Dadras:2017DETC,Taghavian:2019AJC}, controller synthesis \cite{Yin:2017AMM,Nemati:2019AJC}, and optimization algorithm \cite{Chen:2017AMC,Yin:2018AMM}, etc. Due to the great efforts devoted by researchers, a large number of valuable results have been reported on fractional calculus. For the details of the most recent advances, one can refer to some excellent papers \cite{Chen:2018NODY,Xue:2017FCAA,Sun:2018CNSNS} and the references therein.

Despite fractional calculus is the extension of the traditional calculus, it has complex definition which brings it long memory characteristic \cite{Wei:2017FCAAb}. Therefore, it is difficult to simulate fractional order systems and implement fractional order controllers \cite{Yin:2018IS}. From another perspective, fractional order systems are essentially infinite dimensional \cite{Trigeassou:2013CMA}, which makes it difficult to obtain their analytical time--domain response especially for the nonlinear case under nonzero initial conditions. Thus, numerical approximation problem comes into being and becomes essential and challenging. Fundamentally speaking, the core task of numerical approximation is to approximate the fractional calculus operators. Following this idea, many valuable results have been produced in the solution of continuous time fractional order systems \cite{Oustaloup:2000TCSI,Poinot:2003SP,Wei:2014IJCAS,Liang:2014IJSS,Wei:2019ISA,Wei:2016ISA,Du:2017IJCAS}.

It is worth mentioning that the most famous work in numerical approximation is the Oustaloup method \cite{Oustaloup:2000TCSI}. The main idea of this method is to use a series of polylines with the slope of $20{\rm dB/dec}$ and $0{\rm dB/dec}$ to approximate the magnitude--frequency curve that is an elegant straight line with the slope of $20\alpha{\rm dB/dec}, \alpha\in(0,1)$. However, since the designed poles will be different with different differential orders, the dimension of the approximate model with multiple differential operator will be very large, which is not conducive to practical use. To reduce the dimension of the approximation model, the so-called fixed-pole approximation method was proposed in \cite{Liang:2014IJSS,Wei:2019ISA}. In this method, the poles will be fixed for each integral order $\alpha\in(0,1)$, which will bring a smaller dimensional approximate model without loss of the precision. Although the aforementioned methods have been improved, all of them are still based on curve fitting method. Since the actual curve is not a broken line and the poles are limited to real numbers, these approximation methods are flawed.
Bearing this in mind, the frequency--domain identification based method was considered \cite{Wei:2016ISA,Du:2017IJCAS}. However, all of them focused on the continuous time system and none of the discrete time system was considered.

With the rapid development and wide application of digital computers, discrete--time systems become increasingly popular. In this process, some pioneering work have been done on discrete fractional calculus and discrete fractional order systems \cite{Cheng:2011Book,Ostalczyk:2015Book,Goodrich:2015Book,Wu:2017AMC,Wei:2019CNSNSa}. Similarly, the numerical implementation of discrete fractional order systems is a thorny and task. To solve this problem, two discretization schemes were presented for continuous fractional differentiators and integrators \cite{Chen:2002TCSI}. Maione proved that the Laguerre approximation to the Tustin fractional operator $s^{-\alpha}$ was stable and minimum-phase \cite{Maione:2013TAC}. Fractional order Euler-Frobenius polynomials was defined to characterize the asymptotic sampling zeros for fractional systems as the sampling period tends to zero \cite{Yucra:2013TAC}. An original direct discretization method was designed to produce low integer order $z$ domain transfer functions via inverse fast Fourier transform \cite{De:2018ISA}. 
Different from the discretization of continuous time fractional order systems, the numerical scheme in the matrix form is given for discrete fractional order systems directly \cite{Sierociuk:2016JVC}. However, this method generally does not perform effectively when the memory length changes and higher computational complexity will be brought. By means of the singular value decomposition-originated balanced truncation method, an approach was proposed to approximate the linear time invariant discrete time fractional order systems \cite{Stanis:2017JFI}. However, it is not convenient to select the interested frequency range and the approximate degree by this approach. With further research, it has found that this issue is fundamentally an approximation problem of fractional sum operation $\nabla^{-\alpha}$ in the frequency domain of nabla Laplace transform. Inspired by this, the purpose of this paper is to develops a new strategy that not only approximates fractional order systems with high accuracy, but also differs from traditional curve fitting methods. To complete this, there are many challenges. i) It is difficult to realize a system in the frequency domain. ii) The nonlinear parameter identification problem hinders the access of approximation model. iii) The conditions on nonzero initial instant and nonzero initial state bring more difficulty further.

Before end up this section, the main contributions of this paper are summarized as follows.

\begin{enumerate}[i)]
\item The inverse nabla Laplace transform is defined and proven for the first time, which implies the nonzero initial instant problem. Afterwards, the method to calculate $f(k)$ from its nabla Laplace transform $F(s)$ is discussed in detail.
\item Inspired by the unilateral frequency distributed model, three bilateral equivalent frequency distributed models are derived for fractional sum equation. Afterwards, the equivalent frequency distributed models for three special fractional difference equations are deduced.
\item Based on the infinite dimensional model of nabla fractional dynamic system, the approximation problem is transformed into a nonlinear parameter identification problem in frequency domain, and the needed parameters are determined by vector fitting method \cite{Gustavsen:1999TPD}.
\item The proposed approximation strategy is applied to simulate nabla fractional dynamic systems with nonzero initial instant and nonzero initial state. Though this method is developed from the frequency domain, it could also be applied for the nonlinear case.
\end{enumerate}

The remainder of this paper is organized as follows. Section II presents preliminaries, including some fundamental knowledge of nabla discrete fractional calculus and nabla Laplace transform. Section III proposes the numerical approximation method for fractional sum operators and nabla fractional order systems. To illustrate the validity of proposed approach, three illustrative examples are contained in Section IV. Finally, some conclusions are drawn in Section V.


\section {Preliminaries}\label{Section 2}
In this section, the fundamental knowledge for nabla fractional calculus and the preparatory problem statement \cite{Goodrich:2015Book} will be provided.

The $\alpha$-th nabla fractional sum of a function $ f $: $\mathbb{N}_{a+1} \rightarrow\mathbb{ R}$ can be defined by
\begin{equation}\label{Eq1}
{\textstyle {}_a^{}\nabla_k^{-\alpha}f(k)\triangleq \sum\nolimits_{i=0}^{k-a-1}(-1)^{i} \bigl(\begin{smallmatrix}-\alpha\\ i \end{smallmatrix}\bigr) f(k-i),}
\end{equation}
where $\alpha>0$, $k\in\mathbb{N}_{a+1}$, $\mathbb{N}_{a+1}\triangleq\{a+1,a+2,a+3,\cdots\}$, $\bigl(\begin{smallmatrix}p\\ q\end{smallmatrix}\bigr)=\frac{{\rm{\Gamma}}(p+1)}{{\rm{\Gamma}}(q+1){\rm{\Gamma}}(p-q+1)}$ and ${\rm \Gamma}(\cdot)$ is the Gamma function.

With this definition, the nabla Caputo fractional difference is defined by
\begin{equation}\label{Eq2}
{{}_a^{}\nabla_k ^\alpha }f\left( k \right) \triangleq {{}_a^{}\nabla_k ^{ \alpha-n }}{\nabla ^n}f\left( k \right),
\end{equation}
where $n-1< \alpha<n$, $n\in\mathbb{N}_+$, and ${\nabla ^n}$ represents the normal $n$-th backward difference ${\nabla ^n}f\left( k \right) \triangleq \sum\nolimits_{i = 0}^n {{{{\left(\hspace{-1pt}{ - 1} \hspace{-1pt}\right)}^i}\left({\begin{smallmatrix}
{ n }\\
i
\end{smallmatrix}}\right)}f\left( {k - i} \right)}$.

The nabla Laplace transform  of a function $ f $: $\mathbb{N}_{a+1} \rightarrow\mathbb{ R}$ is defined as \cite{Atici:2009EJQTDE}
\begin{equation}\label{Eq3}
{\textstyle \mathscr{N}_a\{f(k)\} \triangleq \sum\nolimits_{k=1}^{+\infty}(1-s)^{k-1}f(k+a), s\in \mathbb{C}.}
\end{equation}
Due to the existence of the parameter $a$, nonzero initial instant problem can be solved accordingly. Assuming $F(s) =\mathscr{N}_a\{f(k)\}$, then one has \cite{Wei:2018arXiv}
\begin{equation}\label{Eq4}
{\textstyle f\left( {a + \kappa } \right) = \mathop {\lim }\limits_{s \to 1} \frac{{F\left( s \right) - \sum\nolimits_{k = 1}^{\kappa  - 1} {{{\left( {1 - s} \right)}^{k - 1}}f\left( {a + k} \right)} }}{{{{\left( {1 - s} \right)}^{\kappa  - 1}}}},}
\end{equation}
where $\kappa\in\mathbb{Z}_+$. Till now, it can be concluded that given $F(s)$, the value of $f(k),~k\in\mathbb{N}_{a+1}$ can be computed. However, as $k$ increases gradually, calculating the value of $f(k)$ one by one has proved difficult. Taking limit operation is not an easy task either. As a consequence, an alternative solution is expected.

\vspace{-10pt}
\begin{theorem}\label{Theorem 1}
If $F(s)={{\mathscr N}_a}\left\{ {f(k)} \right\}$ with $f:\mathbb{N}_{a+1} \to \mathbb{R}$ and  $a\in\mathbb{R}$, then the inverse nabla Laplace transform can be expressed as
\begin{equation}\label{Eq5}
{\textstyle
\begin{array}{rl}
f\left( k \right) =&\hspace{-6pt} {\mathscr N}_a^{ - 1}\{ F\left( s \right)\} \\
 \triangleq&\hspace{-6pt} \frac{1}{{2\pi \rm{j}}}\oint_c {F\left( s \right){{(1 - s)}^{ - k + a}}{\rm{d}}s} ,k \in {\mathbb{N}_{a + 1}},
 \end{array}}
\end{equation}
 where $c$ is a closed curve rotating around the point $(1,{\rm j}0)$ clockwise and it also locates in the convergent domain of $F(s)$.
\end{theorem}
\noindent
\textbf{Proof}
For the given curve $c$, then one has
\begin{eqnarray}\label{Eq6}
{\textstyle \begin{array}{l}
\frac{1}{{2\pi {\rm{j}}}}\oint_c {F\left( s \right){{(1 - s)}^{ - k + a}}{\rm{d}}s} \\
 = \frac{1}{{2\pi {\rm{j}}}}\oint_c {\sum\nolimits_{i = 1}^{ + \infty } {{{(1 - s)}^{i - 1}}f(i + a)} {{(1 - s)}^{ - k + a}}{\rm{d}}s} \\
 = \frac{1}{{2\pi {\rm{j}}}}\sum\nolimits_{i = 1}^{ + \infty } {f(i + a)} \oint_c {{{(1 - s)}^{i - k + a - 1}}{\rm{d}}s}.
\end{array}}
\end{eqnarray}

Define $s = 1 - r{{\rm e}^{-{\rm{j}}\theta }}$ with $r > 0$, $\theta\in(-\pi,\pi)$ and keep the integration curve $c$ is inside the convergent region of $F\left( s \right)$. Then, ${\rm{d}}s = {\rm{j}}r{{\rm e}^{-{\rm{j}}\theta }}{\rm{d}}\theta $ and
\begin{eqnarray}\label{Eq7}
{\textstyle \begin{array}{l}
\frac{1}{{2\pi {\rm{j}}}}\oint_c {F\left( s \right){{(1 - s)}^{ - k + a}}{\rm{d}}s} \\
 = \frac{1}{{2\pi {\rm{j}}}}\sum\nolimits_{i = 1}^{ + \infty } {f(i + a)} \int_{ - \pi }^\pi  {{{(r{{\rm e}^{-{\rm{j}}\theta }})}^{i - k + a - 1}}{\rm{j}}r{{\rm e}^{-{\rm{j}}\theta }}{\rm{d}}\theta } \\
 = \frac{1}{{2\pi }}\sum\nolimits_{i = 1}^{ + \infty } {{r^{i - k + a}}}{f(i + a)} \int_{ - \pi }^\pi  {{{\rm e}^{-{\rm{j}}\theta\left( {i - k + a} \right) }}{\rm{d}}\theta }.
\end{array}}
\end{eqnarray}

With the help of the following fact
\begin{equation}\label{Eq8}
{\textstyle \int_{ - \pi }^\pi  {{{\rm e}^{-{\rm{j}}\theta\left( {i - k + a} \right) }}{\rm{d}}\theta }  = \left\{ \begin{array}{rl}
2\pi &,i = k - a,\\
0&,i \ne k - a,
\end{array} \right.}
\end{equation}
one has
\begin{eqnarray}\label{Eq9}
{\textstyle \begin{array}{rl}
\frac{1}{{2\pi {\rm{j}}}}\oint_c {F\left( s \right){{(1 - s)}^{ - k + a}}{\rm{d}}s} =&\hspace{-6pt} \frac{{{r^0}}}{{2\pi }}f(k - a + a)2\pi \\
=&\hspace{-6pt} f(k).
\end{array}}\hspace{-6pt}
\end{eqnarray}
From the existence and uniqueness of nabla Laplace transform, the inverse nabla Laplace transform can be expressed as
\begin{eqnarray}\label{Eq10}
{\textstyle
{\mathscr N}_a^{ -1}\{ F\left( s \right)\} =  \frac{1}{{2\pi {\rm{j}}}}\oint_c {F\left( s \right){{(1 - s)}^{ - k + a}}{\rm{d}}s} ,}
\end{eqnarray}
where $k \in {\mathbb{N}_{a + 1}}$. This completes the proof. \hfill $\Box$

It is noteworthy that if the nabla Laplace transform $F(s)$ is given, $f(k)$ can be calculated via Theorem \ref{Theorem 1}. Sometimes, it is difficult to solve such a contour integral problem. Especially, in many situations, $F(s)$ is not given previous or impossible to access directly. For example, $f(k)$ is the state or output of a nonlinear system. Thus, a more effective and practical approach is to be proposed.

With the help of the frequency distribution model theory in \cite{Wei:2019CNSNSb}, the system ${}_a^{ }\nabla_k^{-\alpha}u(k) = y(k)$ with $0<\alpha<1$ can be expressed equivalently as the following unilateral frequency distributed models
\begin{equation}\label{Eq11}
{\textstyle\left\{
\begin{array}{rl}
\nabla z(\omega, k)=&\hspace{-6pt} -\omega z(\omega, k) + u(k),\\
y(k)=&\hspace{-6pt}  \int _0^{+\infty} \mu_\alpha(\omega) z(\omega, k){\rm d}\omega,
\end{array}
\right.  }
\end{equation}
\begin{equation}\label{Eq12}
{\textstyle\left\{
\begin{array}{rl}
\nabla z(\omega, k)=&\hspace{-6pt} -\omega z(\omega, k) + \mu_\alpha(\omega) u(k),\\
y(k)=&\hspace{-6pt}  \int _0^{+\infty} z(\omega, k){\rm d}\omega,
\end{array}
\right.  }
\end{equation}
\begin{equation}\label{Eq13}
{\textstyle\left\{
\begin{array}{rl}
\nabla z(\omega, k)=&\hspace{-6pt} -\omega z(\omega, k) + \sqrt{\mu_\alpha(\omega)}u(k),\\
y(k)=&\hspace{-6pt}  \int _0^{+\infty} \sqrt{\mu_\alpha(\omega)} z(\omega, k){\rm d}\omega,
\end{array}
\right.  }
\end{equation}
where $u(k)$ is the input, $y(k)$ is the output, $z\left( \omega,k \right)$ is the state, ${\mu _\alpha }\left( \omega  \right)=\frac{{\sin \left( {\alpha \pi } \right)}}{{{\omega ^{\alpha} }\pi }}$ is the weight function and $z\left( \omega,a \right)=0$ is the initial value. It is noted that different locations of the weight function give different models, such as, the output form in (\ref{Eq11}), the input form in (\ref{Eq12}) and the balance form in (\ref{Eq13}).

Likewise, different equivalent bilateral frequency distributed model of ${}_a^{ }\nabla_k^{-\alpha}u(k) = y(k)$ with $0<\alpha<1$ can be obtained as follows.
\begin{equation}\label{Eq14}
{\textstyle\left\{
\begin{array}{rl}
\nabla z(\omega, k)=&\hspace{-6pt} -\omega^2 z(\omega, k) + u(k),\\
y(k)=&\hspace{-6pt}  \int _{-\infty}^{+\infty} \omega\mu_\alpha(\omega^2)z(\omega, k){\rm d}\omega,
\end{array}
\right.  }
\end{equation}
\begin{equation}\label{Eq15}
{\textstyle\left\{
\begin{array}{rl}
\nabla z(\omega, k)=&\hspace{-6pt} -\omega^2 z(\omega, k) + \mu_\alpha(\omega^2)u(k),\\
y(k)=&\hspace{-6pt}  \int _{-\infty}^{+\infty} \omega z(\omega, k){\rm d}\omega,
\end{array}
\right.  }
\end{equation}
\begin{equation}\label{Eq16}
{\textstyle\left\{
\begin{array}{rl}
\nabla z(\omega, k)=&\hspace{-6pt} -\omega^2 z(\omega, k) + \sqrt{\mu_\alpha(\omega^2)}u(k),\\
y(k)=&\hspace{-6pt}  \int _{-\infty}^{+\infty} \omega \sqrt{\mu_\alpha(\omega^2)}z(\omega, k){\rm d}\omega.
\end{array}
\right.  }
\end{equation}

If the fractional sum equation ${}_a^{}\nabla_k^{-\alpha}u(k) = y(k)$ is replaced by the difference one ${}_a^{}\nabla_k^{\alpha}y(k) = u(k)$, the corresponding equivalent model in the output form can be expressed as
\begin{equation}\label{Eq17}
   {\textstyle \left\{ {\begin{array}{rl}
   {\nabla}z\left( \omega,k \right) =&\hspace{-6pt}  - \omega z\left( \omega,k \right) + u\left( k \right),\\
   y\left( k \right) =&\hspace{-6pt} \int_0^{+\infty}  {{\mu _\alpha }\left(\omega\right)z\left( \omega,k \right){\rm d}\omega} ,
   \end{array}} \right.}
  \end{equation}
where $\alpha\in(0,1)$ and $z\left( \omega,a \right)=\frac{\delta\left( \omega \right)}{{\mu _\alpha }\left( \omega\right)}x\left( a \right)$ which is different from the zero initial value in (\ref{Eq11})-(\ref{Eq16}).

In (\ref{Eq17}),  $u(k)$ is known and $y(k)$ is to be calculated. On the contrary, when $y(k)$ is known and $u(k)$ is to be calculated, then the formula $u(k)={}_a^{ }\nabla_k^{\alpha}y(k)$ can be equivalently expressed as
\begin{equation}\label{Eq18}
   {\textstyle \left\{ {\begin{array}{rl}
   {\nabla}z\left( \omega,k \right) =&\hspace{-6pt}  - \omega z\left( \omega,k \right) + {\nabla}y\left( k \right),\\
   u\left( k \right) =&\hspace{-6pt} \int_0^{+\infty}  {{\mu _{1-\alpha} }\left(\omega\right)z\left( \omega,k \right){\rm d}\omega} ,
   \end{array}} \right.}
  \end{equation}
where $\alpha\in(0,1)$ and $z\left( \omega,a \right)=0$.

In (\ref{Eq17}), the order $\alpha\in(0,1)$. If $\alpha\in(n-1,n)$, $n\in\mathbb{Z}_+$, the system ${}_a^{ }\nabla_k^{\alpha}y(k) = u(k)$ can be equivalently expressed as
  \begin{eqnarray}\label{Eq19}
   \left\{ {\begin{array}{rl}
   {\nabla}z\left( \omega,k \right) =&\hspace{-6pt}  - \omega z\left( \omega,k \right) + u\left( k \right),\\
   \sigma\left( k \right) =&\hspace{-6pt} \int_0^{+\infty}  {{\mu _{\alpha-n+1} }\left(\omega\right)z\left( \omega,k \right){\rm d}\omega} ,\\
   {\nabla^{n-1}}y\left( k \right)=&\hspace{-6pt}\sigma\left( k \right),
   \end{array}} \right.
  \end{eqnarray}
where $z\left( \omega,a \right)=\frac{\delta\left( \omega \right)}{{\mu _\alpha }\left( \omega\right)}\nabla^{n-1} x\left( a \right)$.

Note that all the developed frequency distributed models lay the groundwork for the numerical implementation for nabla fractional order systems. They are infinite dimensional and cannot be used directly. However, their transfer functions are derived from the following identical equation \cite{Wei:2016ISA}
  \begin{equation}\label{Eq20}
   {\textstyle \frac{1}{s^\alpha} = \int_{ 0 }^{ + \infty } {\frac{{{\mu _\alpha }( \omega  )}}{{s + \omega }}{\rm{d}}\omega },}
  \end{equation}
with $\alpha\in(0,1)$, $s\in\mathbb{C}\backslash\mathbb{R}_-$, which provides the possible to solve the numerical simulation problem.

\section{Numerical Approximation Scheme}\label{Section 3}
In this section, an effective algorithm is developed to approximate the nabla fractional sum operator $S_\alpha (s)=\frac{1}{s^\alpha}$ via system identification technique. To deal with the nonlinear coupling problem, the vector fitting method is introduced. From this, the numerical approximation of nabla fractional order systems is further investigated.

\subsection{For nabla fractional sum operator}

To simulate the nabla fractional order system with infinite dimensional characteristic, a practical solution is to discretize the continuous frequency range $\omega \in[0, +\infty)$ with finite distributed frequency points $\omega_0,\omega_1,\cdots,\omega_N$. From this, an approximate finite dimensional state space model can be expressed as
\begin{equation}\label{Eq21}
{\textstyle    \left\{
        \begin{array}{rl}
\nabla z(\omega_i, k)=&\hspace{-6pt}  -\omega_iz(\omega_i, k) + u(k),\\
y(k)=&\hspace{-6pt}  \sum\nolimits_{i=0}^{N}c_iz(\omega_i, k),
        \end{array}
    \right. }
\end{equation}
where $ c_i = \mu_\alpha(\omega_i)( \omega_{i} - \omega_{i-1})$ is the weight value and $z(\omega_i, k)\in\mathbb{R}$ is the approximate system state. Similarly, the transfer function of the above model (\ref{Eq21}) can be obtained as follows
\begin{equation}\label{Eq22}
{\textstyle \hat{S}_\alpha (s) = \sum\nolimits_{i=0}^{N}\frac{c_i}{s+\omega_i}=\frac{G_\alpha}{s+\omega_0}\prod\nolimits_{i = 1}^N {\frac{{s + {{\bar \omega }_i}}}{{s + {\omega _i}}}} .}
\end{equation}
Therefore, the task of numerical approximation is to approximate $S_\alpha (s)$ with $\hat{S}_\alpha (s)$. It has been pointed by the reference  \cite{Montseny:1998PFDS} that if the conditions that $N\to+\infty$, $\omega_0 \to 0$, $\omega_N \to +\infty$, $\mathop {\sup }\limits_{1 \le i \le N} \left| {{\omega _i} - {\omega _{i - 1}}} \right| \to 0$ are satisfied, then the system (\ref{Eq22}) would approximate the nabla fractional sum operator $S_\alpha (s)$ with arbitrary accuracy. In this case, the finite dimensional frequency distribution model (\ref{Eq21}) can be adopted to approximate the infinite dimensional frequency distribution model (\ref{Eq11}). Therefore, the problem of operator approximation has been transformed into a system identification problem: when the order $\alpha$ and $s_l, l = 1, 2, \cdots, L\in\mathbb{Z}_+$ are known, how to determine the parameters $\omega_i$ and $c_i, i= 0, 1,\cdots, N\in\mathbb{N}$, so as to minimize the error between $S_\alpha (s)$ and $\hat{S}_\alpha (s)$, that is
\begin{equation}\label{Eq23}
{\textstyle {\rm arg}\hspace{1mm} \min\limits_{c_i, \omega_i}\sum\nolimits_{l=1}^{L}\big|\frac{1} {s_l^\alpha}-\sum\nolimits_{i=0}^{N}\frac{c_i}{s_l+\omega_i} \big|^{2}.}
\end{equation}

At superficial glance, this is a nonlinear identification problem. To solve this problem, it is necessary to introduce two auxiliary transfer functions as
\begin{equation}\label{Eq24}
{\textstyle\left\{
\begin{array}{rl}
h(s)&\hspace{-6pt} \triangleq\prod\nolimits_{i = 0}^N {\frac{{s + {{\omega }_i}}}{{s + {p _i}}}} = \sum\nolimits_{i=0}^{N}\frac{\lambda_i}{s+p_i}+1,\\
H(s)&\hspace{-6pt}\triangleq\frac{1}{s+p_0}\prod\nolimits_{i = 1}^N {\frac{{s + {\bar \omega_i}}}{{s + {p_i}}}}= \sum\nolimits_{i=0}^{N}\frac{\mu_i}{s+p_i} .
\end{array}
\right.}
\end{equation}
Compared with (\ref{Eq22}) and (\ref{Eq24}), it can be found that the numerator of $h(s)$ is the denominator of $\hat S_\alpha(s)$, and the numerator of $H(s)$ is the numerator of $\hat S_\alpha(s)$. Therefore, as long as the numerators of two auxiliary functions are determined, $\hat S_\alpha(s)$ would be obtained. From this, one can obtain that
\begin{equation}\label{Eq25}
{\textstyle \hat S_\alpha (s) = \frac{H(s)}{h(s)} .}
\end{equation}

For the purpose of obtaining the parameters $\omega_i$ and $c_i$, the following equation should be guaranteed
\begin{equation}\label{Eq26}
{\textstyle S_\alpha (s) = \frac{H(s)}{h(s)} = \frac{\sum\nolimits_{i=0}^{N}\frac{\mu_i}{s+p_i}}{ \sum\nolimits_{i=0}^{N}\frac{\lambda_i}{s+p_i}+1},}
\end{equation}
which can be easily expanded as
\begin{equation}\label{Eq27}
{\textstyle\sum\nolimits_{i=0}^{N}\frac{\mu_i}{s+p_i}-\big( \sum\nolimits_{i=0}^{N}\frac{\lambda_i}{s+p_i} \big)S_\alpha (s)=S_\alpha (s),}
\end{equation}
where $p_k$ is the known parameters when introducing the auxiliary transfer functions. In general, we choose the given discrete frequency points $s_l = {\rm j\,} \zeta_l$ with $ \zeta_l\in \mathbb{R}_+$, $l = 1, 2, \cdots, L$. Then, $S_\alpha (s)$ can be regarded as the frequency--domain response. Obviously, the unknown parameters in (\ref{Eq27}) are only $\mu_i$ and $\lambda_i$ which only exist in the numerators. Thus, the original nonlinear least squares problem in (\ref{Eq23}) is successfully transformed into the following problem
\begin{equation}\label{Eq28}
y(s_l) = \phi^{\rm{T}}(s_l)\theta,
\end{equation}
whose least square solution satisfies
\begin{equation}\label{Eq29}
{\textstyle \min\limits_\theta \sum\nolimits_{l=1}^{L}\left[y(s_l)-\phi(s_l)\theta\right]^2,}
\end{equation}
where
\[\begin{array}{l}
y({s_l}) = {[{\mathop{\rm Re}\nolimits} \left\{ {{S_\alpha }({s_l})} \right\},{\mathop{\rm Im}\nolimits} \left\{ {{S_\alpha }({s_l})} \right\}]^{\rm{T}}},\\
\theta  = {\left[ {{\mu _0}, \ldots ,{\mu _N},{\lambda _0}, \ldots ,{\lambda _N}} \right]^{\rm{T}}}, \\
\phi ({s_l}) =\left[ {{\mathop{\rm Re}\nolimits} \left\{ {\varphi ({s_l})} \right\},{\mathop{\rm Im}\nolimits} \left\{ {\varphi ({s_l})} \right\}} \right],\\
\varphi ({s_l}) = \big[ {\frac{1}{{{s_l} + {p_0}}}, \ldots ,\frac{1}{{{s_l} + {p_N}}},\frac{{ - {S_\alpha }({s_l})}}{{{s_l} + {p_0}}}, \ldots ,\frac{{ - {S_\alpha }({s_l})}}{{{s_l} + {p_N}}}} \big]^{\rm{T}}.
\end{array}\]

Taking what is the above mentioned into account, the estimated values of $\mu_k$ and $\lambda_k$ are acquired as
\begin{equation}\label{Eq30}
\hat{\theta}=\left[{\Phi(s)}\Phi^{\rm T}(s)\right]^{-1}{\Phi(s)}Y(s),
\end{equation}
where $\Phi(s)=\left[{\phi(s_1)}, {\phi(s_2)},\hspace{-2pt}\cdots\hspace{-2pt}, {\phi(s_L)}\right]$,
$Y(s) = \left[y(s_1),\right.$  $\left.y(s_2),\cdots,y(s_L)\right]^{\rm T}.$

Note that $S_\alpha(s)$ is an irrational transfer function with infinite degree while $\hat S_\alpha(s)$ is a rational transfer function with finite degree $N+1$. As a result, only a least square solution can be obtained if we want to approximate $S_\alpha(s)$ with $\hat S_\alpha(s)$. Because $p_i$ will affect the bandwidth of $h(s)$ and $H(s)$, the desired case is that $p_i=\omega_i$. It can be observed from (\ref{Eq24}) that when the initial poles $-p_i$ of the auxiliary function are chosen as the exact poles $-\omega_i$ one has $h(s)=1$, $H(s)=S_\alpha(s)$. In this case, the best approximate performance can be achieved. However, one could not have any prior information when choosing auxiliary functions, so it is often difficult to obtain accurate results in one calculation. Fortunately, formula (\ref{Eq24}) also suggests that after the above calculation, the zeros $-\omega_i$ of $h(s)$ can be used as the poles $-p_i$ in the next cycle. In accordance with this iterative relationship, more accurate results would be obtained.

\begin{remark}\label{Remark 1}
It should be noted that the backwards difference $\nabla z(\omega,k)=z(\omega,k)-z(\omega,k-1)$. The approximate model (\ref{Eq21}) cannot be transformed into the existing case $\left\{ \begin{array}{rl}
x\left( {k + 1} \right) =& \hspace{-6pt}Ax\left( k \right) + Bu(k)\\
y(k) =&\hspace{-6pt}  Cx\left( k \right) + Du(k)
\end{array} \right.$, and therefore it cannot be solved directly by some existing modules in MATLAB. However, the obtained approximate model is causal, stable, minimum-phase, and suitable for a digital implementation.
\end{remark}

\subsection{Treatment of conjugate complex numbers}

Compared with the recursive schemes \cite{Oustaloup:2000TCSI,Poinot:2003SP,Wei:2014IJCAS,Liang:2014IJSS,Wei:2019ISA}, one of the advantages of the proposed approach is that the choice of the poles and residues in the approximation model can be extended to the complex field. However, since the transfer function $S_\alpha(s)$ must be strictly regular, the chosen complex poles must be complex conjugate. Besides, considering that the calculation process of the above method is iterative, the complex poles and residuals obtained during each iteration should be conjugate. Thus, it is essential to make some improvements to the method.

In (\ref{Eq24}), assume that the $(i+1)$-th and $(i+2)$-th poles are the complex conjugate, that is
\begin{equation}\label{Eq31}
\left\{\begin{array}{rl}
p_i=&\hspace{-6pt}{\rm Re}(p_i)+{\rm j\,}{\rm Im}(p_i),\\
p_{i+1}=&\hspace{-6pt}{\rm Re}(p_i)-{\rm j\,}{\rm Im}(p_i).\\
\end{array}\right.
\end{equation}
In this basis, the corresponding residues satisfy the following conjugate relationship
\begin{equation}\label{Eq32}
{\textstyle \left\{\begin{array}{rl}
\mu_i=&\hspace{-6pt}{\rm Re}(\mu_i)+{\rm j\,}{\rm Im}(\mu_i),\\
\mu_{i+1}=&\hspace{-6pt}{\rm Re}(\mu_i)-{\rm j\,}{\rm Im}(\mu_i),\\
\lambda_i=&\hspace{-6pt}{\rm Re}(\lambda_i)+{\rm j\,}{\rm Im}(\lambda_i),\\
\lambda_{i+1}=&\hspace{-6pt}{\rm Re}(\lambda_i)-{\rm j\,}{\rm Im}(\lambda_i).
\end{array}\right.}
\end{equation}

Then, the corresponding elements in vector $\varphi (s_l)$ and $\theta$ can be rewritten as
\begin{equation}\label{Eq33}
{\textstyle \left\{\begin{array}{rl}
\varphi(s_l)_{i+1}=&\hspace{-6pt}\frac{1}{s_l+p_i}+\frac{1}{s_l+p_{i+1}},\\
\varphi(s_l)_{i+2}=&\hspace{-6pt}\frac{\rm j}{s_l+p_i}- \frac{\rm j}{s_l+p_{i+1}},\\
\varphi(s_l)_{2i+2}=&\hspace{-6pt}\frac{-S_\alpha(s_l)}{s_l+p_i}+\frac{-S_\alpha(s_l)}{s_l+p_{i+1}},\\
\varphi(s_l)_{2i+3}=&\hspace{-6pt}\frac{-{\rm j}S_\alpha(s_l)}{s_l+p_i}- \frac{-{\rm j}S_\alpha(s_l)}{s_l+p_{i+1}},\\
\end{array}\right.}
\end{equation}
and
\begin{equation}\label{Eq34}
{\textstyle \left\{\begin{array}{rl}
\theta_{i+1}=&\hspace{-6pt}{\rm Re}(\mu_i),\\
\theta_{i+2}=&\hspace{-6pt}{\rm Im}(\mu_i),\\
\theta_{2i+2}=&\hspace{-6pt}{\rm Re}(\lambda_i),\\
\theta_{2i+3}=&\hspace{-6pt}{\rm Im}(\lambda_i).\\
\end{array}\right.}
\end{equation}

In this way, the original least squares solution $\hat\theta$ can be found in the real space and the conjugacy can be guaranteed firmly. The specific computing formula is also (\ref{Eq30}). If the estimated parameter pairs corresponding to $p_i$ and $p_{i +1}$ are $\theta_{i+1},\theta_{i+2}$ and $\theta_{2i+2},\theta_{2i+3}$, then the original complex parameters $(\mu_i, \mu_{i+1})$ and $(\lambda_i, \lambda_{i+1})$ can be recovered via the relationship in equation (\ref{Eq32}) and equation (\ref{Eq34}).

Based on the above operation, the problem with complex poles has been solved successfully. To sum up, the entire calculation process proposed above can be described as Algorithm 1.
\vspace{3mm}
\begin{table*}
\centering
\scriptsize
\begin{tabular}{l}     
\toprule
\textbf{Algorithm 1}: The numerical approximation of fractional sum operator\\        
\midrule
{\bf Prior Information:} the fractional order $\alpha$ and
 the discrete frequency points $s_l,(l=1, 2, \cdots, L)$\\
{\bf Input:} the total iterations number $T$ and the degree of approximation model $N+1$ $(L>2N+2)$\\
{\bf Output:} the poles $-\omega_k$ and the residues $c_i$ $(i=0,1,\cdots,N)$\vspace{2mm}\\
 Step 1: Let the number of iterations $t = 0$ and choose initial poles $-p_i$.\\
 Step 2: Use (\ref{Eq28}) to calculate $y(s_l)$ and $\phi(s_l)$.\\
 Step 3: Use (\ref{Eq30}) to calculate $\hat \theta$, and get $\mu_i$ and $\lambda_i$.\\
 Step 4: Use (\ref{Eq24}) to calculate the zeros $-\omega_i$ of $h(s)$ based on $p_i$ and $\lambda_i$.\\
 Step 5: If $t\geq T$, take the zeros $-\omega_i$ obtained in Step 4 as the poles of ${\hat{S}}_\alpha(s)$\\
 \hspace{23pt}  and use the least squares algorithm to find the residues $c_i$ of ${\hat{S}}_\alpha(s)$. \\
 \hspace{23pt} Otherwise, return to Step 2 and let $p_i=\omega_i,t=t+1$.\\
 Step 6: Finally, put $\omega_i$ and $c_i$ as the parameters of ${\hat{S}}_\alpha(s)$.\\
\bottomrule
\end{tabular}
\end{table*}

\subsection{For nabla fractional dynamic systems}
After approximating the nabla fractional sum operator $\frac{1}{s^\alpha}$, the presented results can be extended to nabla discrete fractional dynamic systems
\begin{equation}\label{Eq35}
\left\{\begin{array}{rl}
{}_a^{}\nabla_k^{{\bf \alpha}} {\bf x}(k)=&\hspace{-6pt} { \bf f}\left({\bf x}(k), { \bf u}(k)\right),\\
{\bf y}(k) =&\hspace{-6pt}   {\bf g}\left({ \bf x}(k), {\bf u}(k)\right),
\end{array}\right.
\end{equation}
where the parameters satisfy
\begin{equation*}
\left\{\begin{array}{l}
\textrm{input}: {\bf u} = \left[u_1, u_2, \cdots, u_p\right]^{\rm T}\in\mathbb{R}^p,\\
\textrm{output}: {\bf y} = \left[y_1, y_2, \cdots, y_q\right]^{\rm T}\in\mathbb{R}^q,\\
\textrm{pseudo state}: {\bf x} = \left[x_1, x_2, \cdots, x_n\right]^{\rm T}\in\mathbb{R}^n,\\
\textrm{fractional order}:  {\bf \alpha} = \left[\alpha_1, \alpha_2, \cdots, \alpha_n\right]^{\rm T}\in\mathbb{R}^n,\\
\textrm{nonlinear/linear function}: {\bf f} = \left[f_1,  f_2, \cdots, f_n\right]^{\rm T},\\
\textrm{nonlinear/linear function}: {\bf g} =\left[g_1, g_2, \cdots, g_q\right]^{\rm T}.
\end{array}\right.
\end{equation*}

By applying the frequency distribution model theory, the finite dimensional pseudo state space system in (\ref{Eq35}) can be rewritten as an infinite dimensional exact state space model
\begin{eqnarray}\label{Eq36}
{\textstyle \left\{\begin{array}{rl}
\nabla {\bf z}(\omega,k)=&\hspace{-6pt} -\omega {\bf z}(\omega,k)+{\bf f}\big({\bf x}(k),{\bf u}(k)\big),\\
{\bf x}(k)=&\hspace{-6pt}\int_0^{+\infty}{\bf \mu}_{{\bf \alpha}}(\omega) {\bf z}(\omega,k){\rm d}\omega,\\
{\bf y}(k)=&\hspace{-6pt}  {\bf g}\big({\bf x}(k),{\bf u}(k)\big),\hspace{-43mm}
\end{array}\right.}
\end{eqnarray}
where $ {\bf z}(\omega, k) = [~z_1(\omega,k),~ z_2(\omega,k),~\cdots,~ z_n(\omega, k)~]^{\rm T}$, ${\bf \mu}_{ \bf\alpha}(\omega) = {\rm diag}\{~\mu_1(\omega),~ \mu_2(\omega),~\cdots,~\mu_n(\omega)~\}$, $\mu_i(\omega) = \frac{\sin(\alpha_i\pi)}{\omega^{\alpha_i}\pi}$ and $z_j\left( \omega,a \right)=\frac{\delta\left( \omega \right)}{{\mu _{\alpha_j} }\left( \omega\right)}x_j\left( a \right)$, $j=1,2,\cdots,n$.

As mentioned earlier, such an infinite dimensional model can not be used directly in practice. Therefore, the system (\ref{Eq35}) needs to be approximated by a finite dimensional model as follows
\begin{eqnarray}\label{Eq37}
{\textstyle \left\{\begin{array}{rl}
\nabla{\bf z}(\omega,k) = &\hspace{-6pt}  M_A {\bf z}(\omega,k) + M_B {\bf f}({\bf x}(k),{\bf u}(k)),\\
{\bf x}(k) = &\hspace{-6pt}M_C{\bf z}(\omega,k),\\
{\bf y}(k) = &\hspace{-6pt}  {\bf g}(\boldsymbol x(k),{\bf u}(k)),
\end{array}\right.}
\end{eqnarray}
where $M_A={\rm diag}\{A_0,A_1,\ldots,A_N\}$, $M_B=[B_0,B_1,$
$\ldots,B_N]^{\rm T}$, $M_C=[C_0,C_1,$ $\ldots,C_N]$, $A_i=-{\rm diag}\{\omega_{1,i},$
$ \omega_{2,i}, \ldots, \omega_{n,i}\}$, $B_i=I_r$, $C_i={\rm diag}\{c_{1,i},c_{2,i},\ldots,c_{n,i}\}$, the parameters $\omega_{j,i}$ and $c_{j,i}$ are generated by the algorithm in previous section for $G_{\alpha_j}(s)$.

The above discussion in (\ref{Eq37}) does not consider the case of nonzero initial state. When  it comes to this situation, just consider how to reasonably assign the initial pseudo state $x(a)$ to the real state $z(\omega,a)$. For the discussed Caputo definition, $z(\omega,a)$ is completely distributed on the frequency point $\omega=0$ and therefore the initial state can be configured as follows
\begin{eqnarray}\label{Eq38}
{\textstyle \left\{
\begin{array}{l}
z_j(\omega_0,a)=\small{\frac{x(a)}{c_{j,0}}},\omega_0=0,j=1,2,\cdots,n,\\
z_j(\omega_i,a)=0,i=1,2,\cdots,N,j=1,2,\cdots,n.
\end{array}
\right.}
\end{eqnarray}
Notably, the value of $\omega_i$ are obtained from identification and $\omega_0=0$ is no longer guaranteed. To achieve this, another approximation model should be assumed i.e., $\hat{S}_\alpha (s) = \frac{c_0}{s}+ \sum\nolimits_{i=1}^{N}\frac{c_i}{s+\omega_i}=\frac{G_\alpha}{s}\prod\nolimits_{i = 1}^N {\frac{{s + {{\bar \omega }_i}}}{{s + {\omega _i}}}}$. In other words, we use ${G_\alpha}\prod\nolimits_{i = 1}^N {\frac{{s + {{\bar \omega }_i}}}{{s + {\omega _i}}}}$ to approximate $s^{1-\alpha}$ and then the zero pole can be guaranteed naturally.

\vspace{-9pt}
\begin{remark}\label{Remark 2}
Just as the previous discussion, the discrete frequency points $s_l$ are selected from the positive imaginary axis. For the continuous time case, the imaginary axis is the boundary to distinguish between stable region and unstable region. This selection strategy will maintain the stability well after approximating. However, for the discrete time case, the imaginary axis is only a curve in stable region and the actual boundary to distinguish between stable region and unstable region is a circle, $(x-1)^2+y^2=1$. Therefore, try to choose $s_l$ in this circle is our ongoing research.
\end{remark}

\begin{remark}\label{Remark 3}
If the adopted Caputo fractional difference is changed by the Riemann--Liouville fractional difference or the Gr\"{u}nwald--Letnikov fractional difference, only the initial value $z\left( \omega,a \right)$ should be replaced. If the nabla case is replaced by the delta case, similar infinite dimensional models can be derived. In this case, the numerical approximation in frequency domain can be performed again.
\end{remark}
\section{Simulation Study}\label{Section 4}
In this section, three numerical examples will be presented to demonstrate the effectiveness of the proposed method.

\vspace{9pt}\noindent\textbf{Example 1: For fractional sum operator}

The range of discrete frequency points is set as $[\omega_l,\omega_h]=[0.001,1000]$ in this section. To illustrate the effectiveness of approximate method, the following approximate error is introduced.
\begin{equation}\label{Eq39}
{\textstyle J = \sum\nolimits_{l=1}^{L}|S_\alpha({\rm j\,}\zeta_l) -\hat{{S}}_\alpha({\rm j\,}\zeta_l)|^2.}
\end{equation}

With different orders $\alpha = 0.1,0.2,\cdots,0.9$, the performance of fractional sum operator approximation is studied, and the results are shown in Table \ref{Table 1}. It can be seen that the approximation error decreases gradually as the order of approximation model $N$ increases.
\begin{table*}[htbp]
    \centering
    \scriptsize
    \caption{The error with different approximation order } \label{Table 1}
    \vspace{6pt}
    \begin{tabular}{cccccccccc}
        \toprule
            $J$& $\alpha = 0.1$ & $\alpha = 0.2$ & $\alpha = 0.3$ & $\alpha = 0.4$ & $\alpha = 0.5$ & $\alpha = 0.6$ & $\alpha = 0.7$ & $\alpha = 0.8$ & $\alpha = 0.9$\\
        \midrule
\hspace{-1.55mm}$N = 5$  &  $ 0.6839$ & $1.2615$ & $1.6254$ & $1.8341$ & $1.9514$ & $1.9669$ & $1.8500$ & $1.5517$ & $1.0079$\\
            $N = 10$ & $ 0.0431$ & $0.0882$ & $0.1409$ & $0.1780$ & $0.2140$ & $0.2725$ & $0.2497$ & $0.3315$ & $0.2095$ \\
            $N = 15$ & $ 0.0090$ & $0.0261$ & $0.0425$ & $0.0997$ & $0.1073$ & $0.1302$ & $0.1868$ & $0.1949$ & $0.1785$\\
            $N = 20$ & $0.0057$ & $0.0407$ & $0.0284$ & $0.0333$ & $0.0610$ & $0.0880$ & $0.1611$ & $0.1360$ & $0.1511$\\
        \bottomrule
    \end{tabular}
\end{table*}

To illustrate the influence of the iteration number further, the approximation for $\frac{1}{s^{0.5}}$ is discussed with results shown in Table \ref{Table 2}. From a macroscopic point of view, as the number of iteration increased, the approximation accuracy also increases. However, from a local point of view, this approach to improve the accuracy is limited. An obvious conclusion is that the better approximation performance can be obtained by determining the number of iterations $T$ around the chosen approximation degree $N+1$. Based on this discovery, it is good enough for the performance of approximation to set the number of iteration in the subsequent simulation as $L = N$, or slightly less than $N$. For convenience, the numbers of iteration in the following examples are set to $T=8$ if not stated.

\begin{table*}[htbp]
    \centering
    \scriptsize
    \caption{The error with different number of iterations} \label{Table 2}
    \vspace{6pt}
    \begin{tabular}{cccccccccc}
        \toprule
            $J$& $T=3$ & $T=6$ & $T=9$ & $T=11$ & $T=12$ & $T=15$ & $T=16$ & $T=18$ & $T=21$\\
        \midrule
\hspace{-1.55mm}$N = 5$   & $10.7382$ & $1.9514$ & $1.9414$ & $2.1431$ & $2.2061$     & $2.2960$     & $2.3080$      & $2.3207$      & $2.3272$\\
            $N = 10$ & $  4.9554$ & $0.2140$ & $0.1003$ & $0.1202$ & $0.1312$     & $0.1549$     & $0.1597$      & $0.1663$      & $0.1712$\\
            $N = 15$ & $ 4.0491$ & $0.1073$ & $0.0105$ & $0.0052$ & $0.0051$      & $0.0069$     & $0.0074$      & $0.0082$      & $0.0088$\\
            $N = 20$ & $ 3.1467$ & $0.0610$ & $0.0053$ & $0.0013$ & $7.73{\rm e}{-4}$ & $2.67{\rm e}{-4}$ & $2.66{\rm e}{-4}$ & $3.23{\rm e}{-4}$ & $4.03{\rm e}{-4}$\\
        \bottomrule
    \end{tabular}
\end{table*}

\begin{figure}[htbp]
  \centering
  \includegraphics[width=0.5\textwidth]{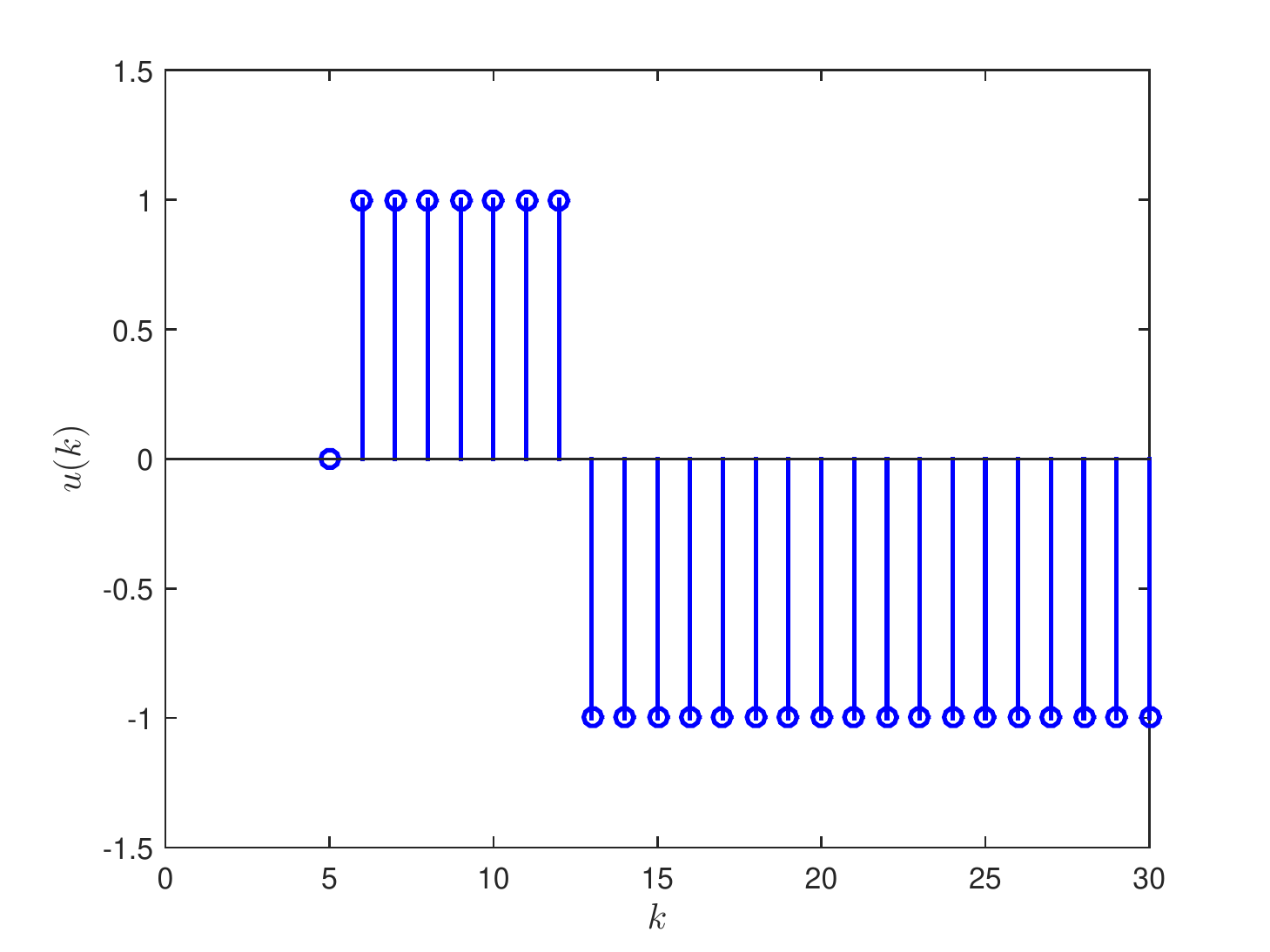}
  \caption{The system input  $u(k)$.}\label{Fig1}
\end{figure}

As can be seen from Table \ref{Table 1} and Table \ref{Table 2}, the approximation accuracy will be improved with the approximate order increased. Actually, the two tables are the comparison between $S_\alpha(s)$ and $\hat{S}_\alpha(s)$. To reflect the approximation performance of the proposed approach more intuitively, the time--domain response is considered. Notably, consider $\alpha=0.5$, $a=5$, $N=20$ and $y(a)=1$ in this and the following examples. Herein, with the following input (see Fig. \ref{Fig1})
\begin{equation}\label{Eq40}
u(k) = \left\{ {\begin{array}{rl}
{0},&{k \le a},\\
{1},&{a<k \le 12},\\
{ - 1},&k>12,
\end{array}} \right.
\end{equation}
one can get the output of system ${}_a^{ }\nabla_k^{-\alpha}u(k) = y(k)$ shown in Fig. \ref{Fig1}.
\begin{figure}[htbp]
  \centering
  \includegraphics[width=0.5\textwidth]{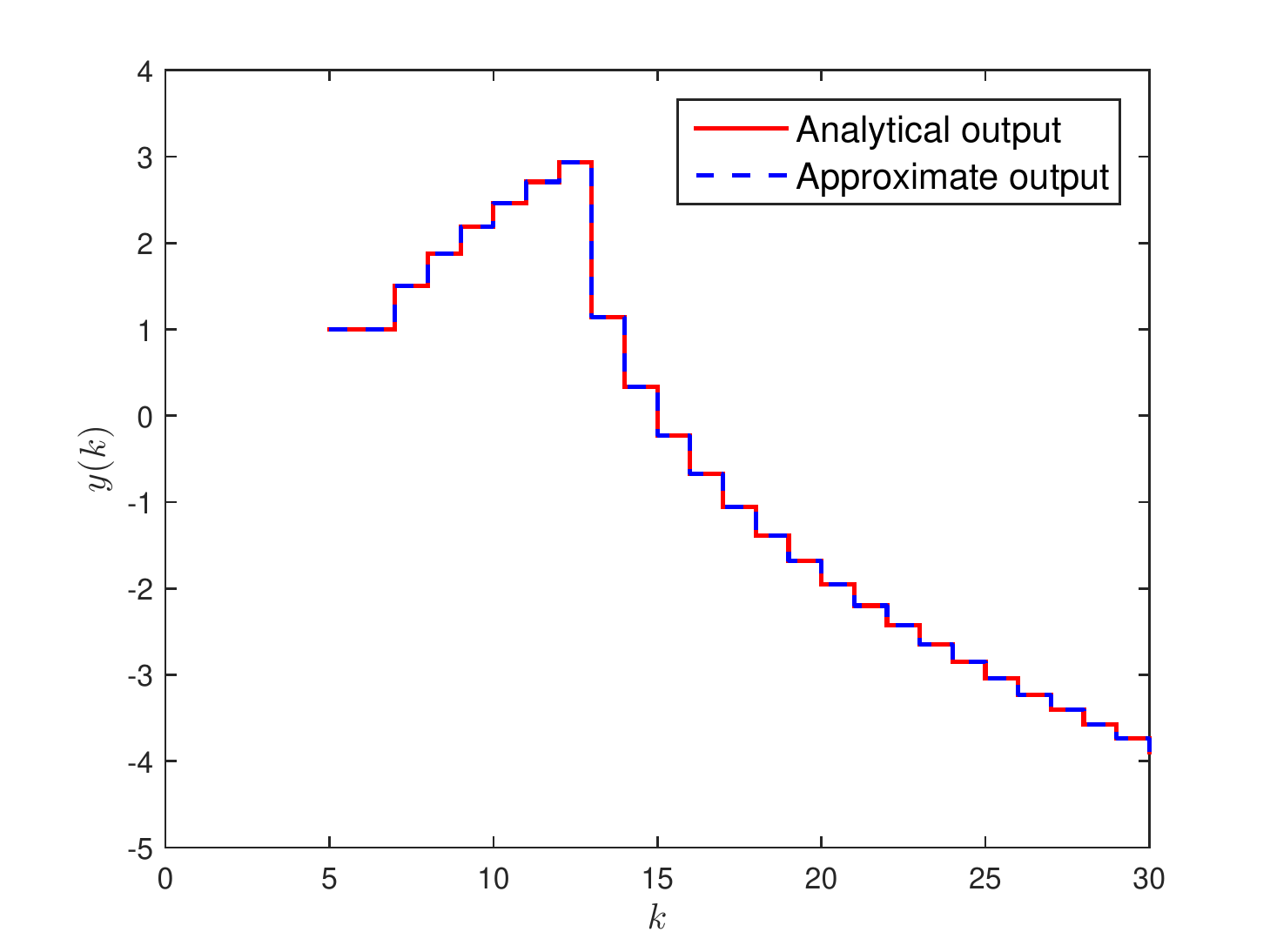}
  \caption{The system output $y(k)$.}\label{Fig2}
\end{figure}

\begin{figure}[htbp]
  \centering
  \includegraphics[width=0.5\textwidth]{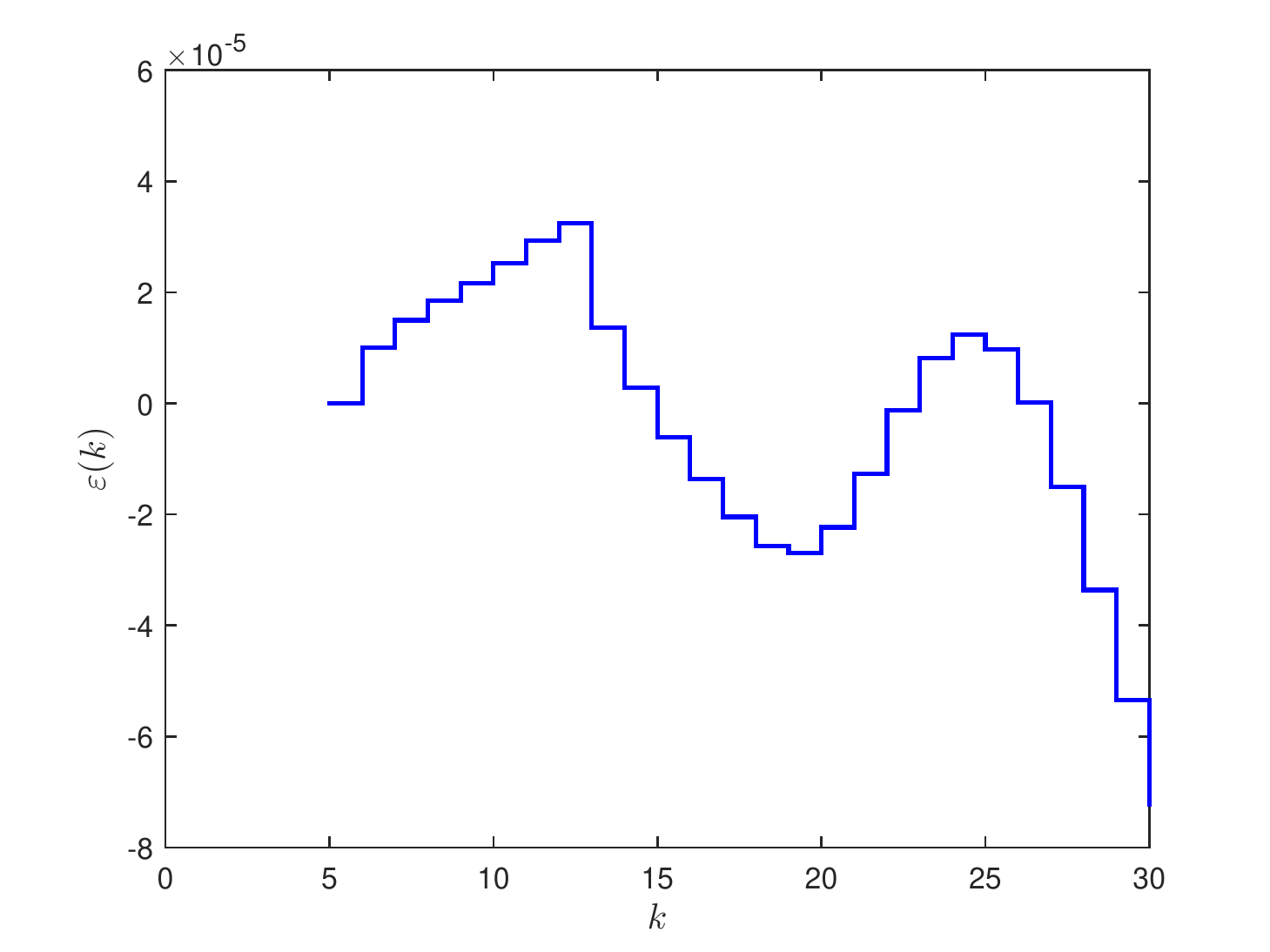}
  \caption{The approximate error $\varepsilon(k)$.}\label{Fig3}
\end{figure}

Fig. \ref{Fig2} illustrates that the results of the proposed method and the analytical solutions are basically the same, and this conclusion can be seen more intuitively from the approximate error $\varepsilon(k)$ in Fig. \ref{Fig3}. The analytical solution of $y(k)$ can be obtained by utilising the definition of fractional sum introduced in Section \ref{Section 2}.

\vspace{24pt}
\noindent\textbf{Example 2: For the linear system}

\begin{figure}[htbp]
  \centering
  \includegraphics[width=0.5\textwidth]{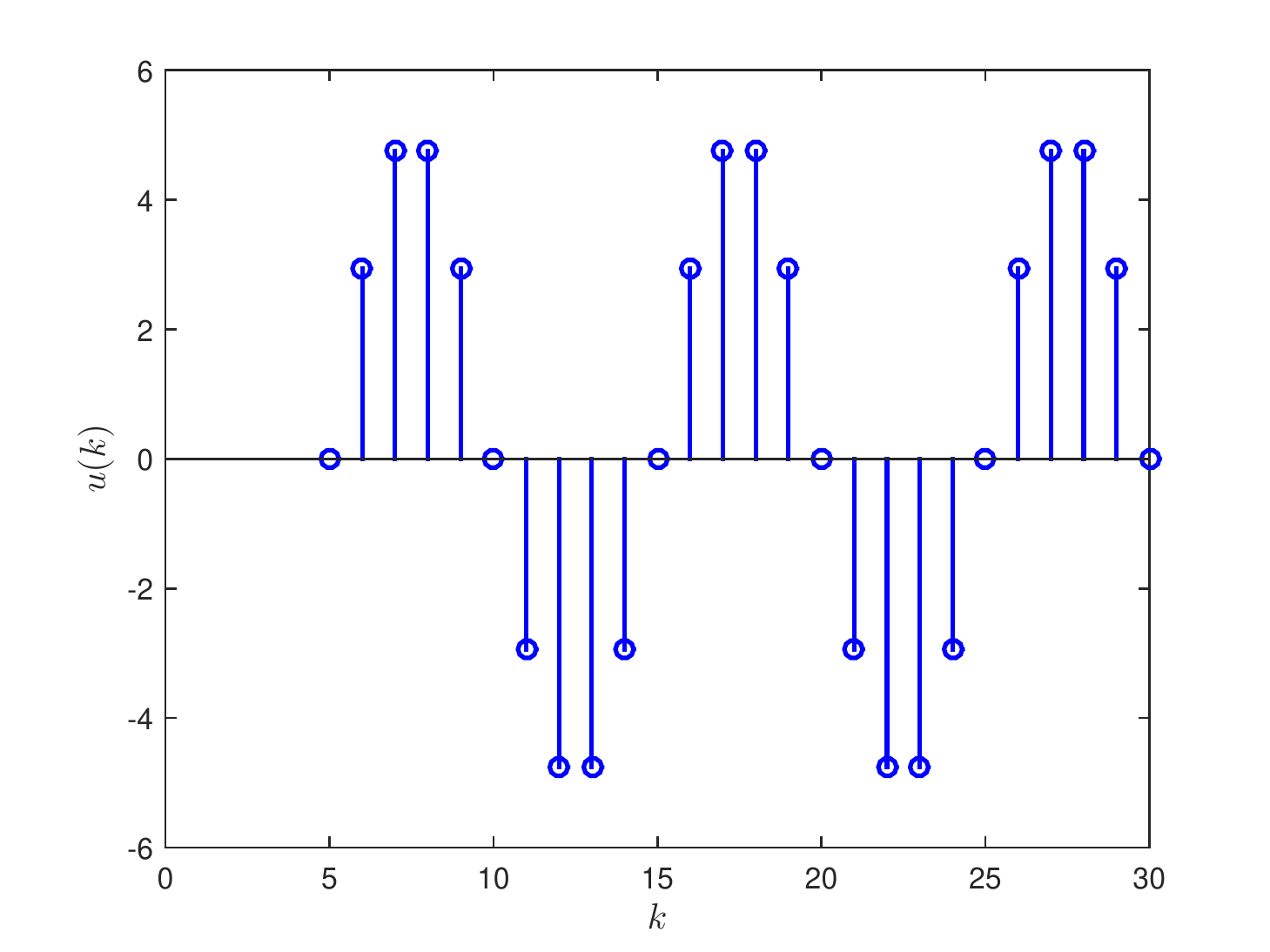}
  \caption{The system input  $u(k)$.}\label{Fig4}
\end{figure}

\begin{figure}[htbp]
  \centering
  \includegraphics[width=0.5\textwidth]{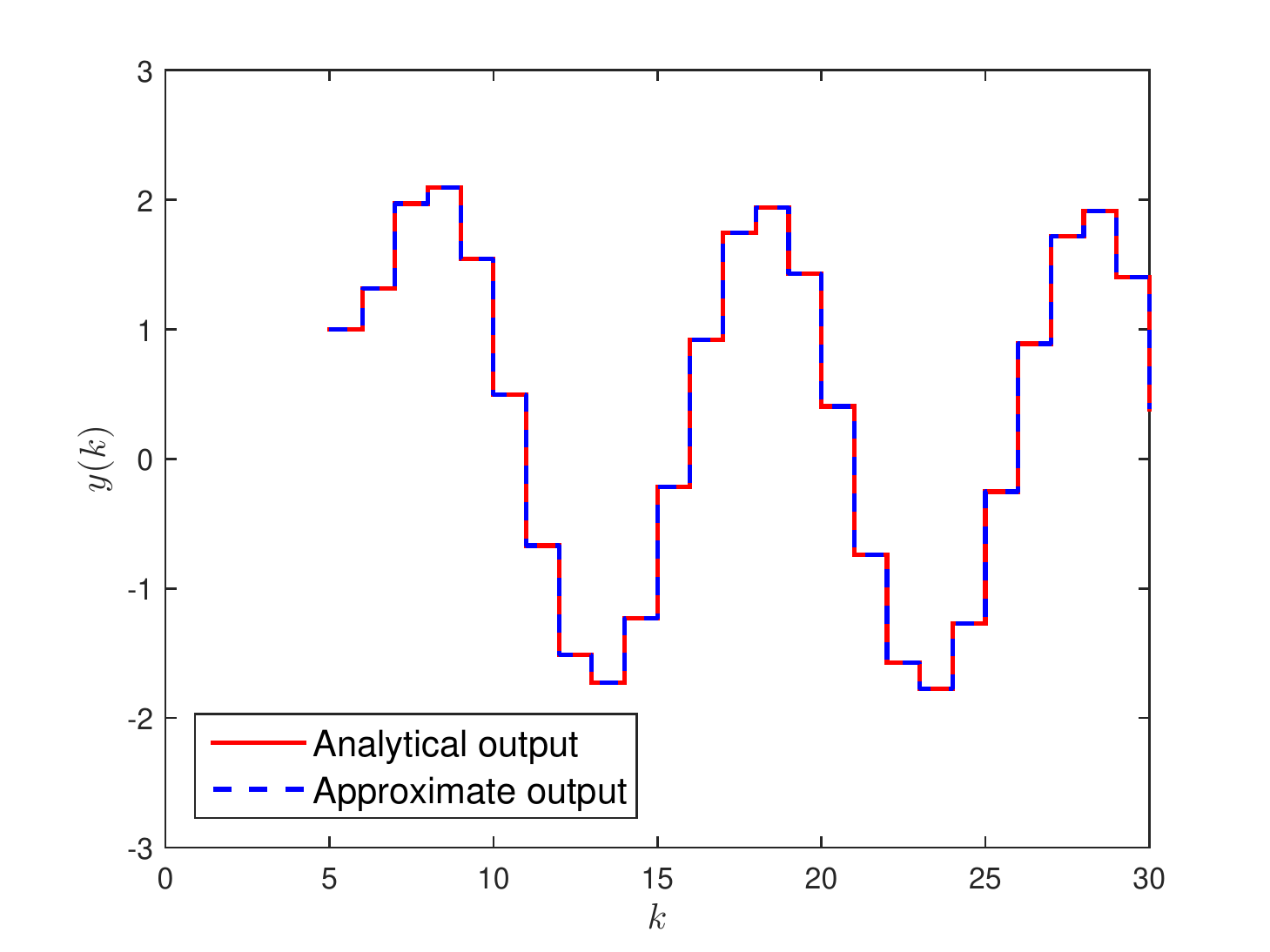}
  \caption{The system output $y(k)$.}\label{Fig5}
\end{figure}

\begin{figure}[htbp]
  \centering
  \includegraphics[width=0.5\textwidth]{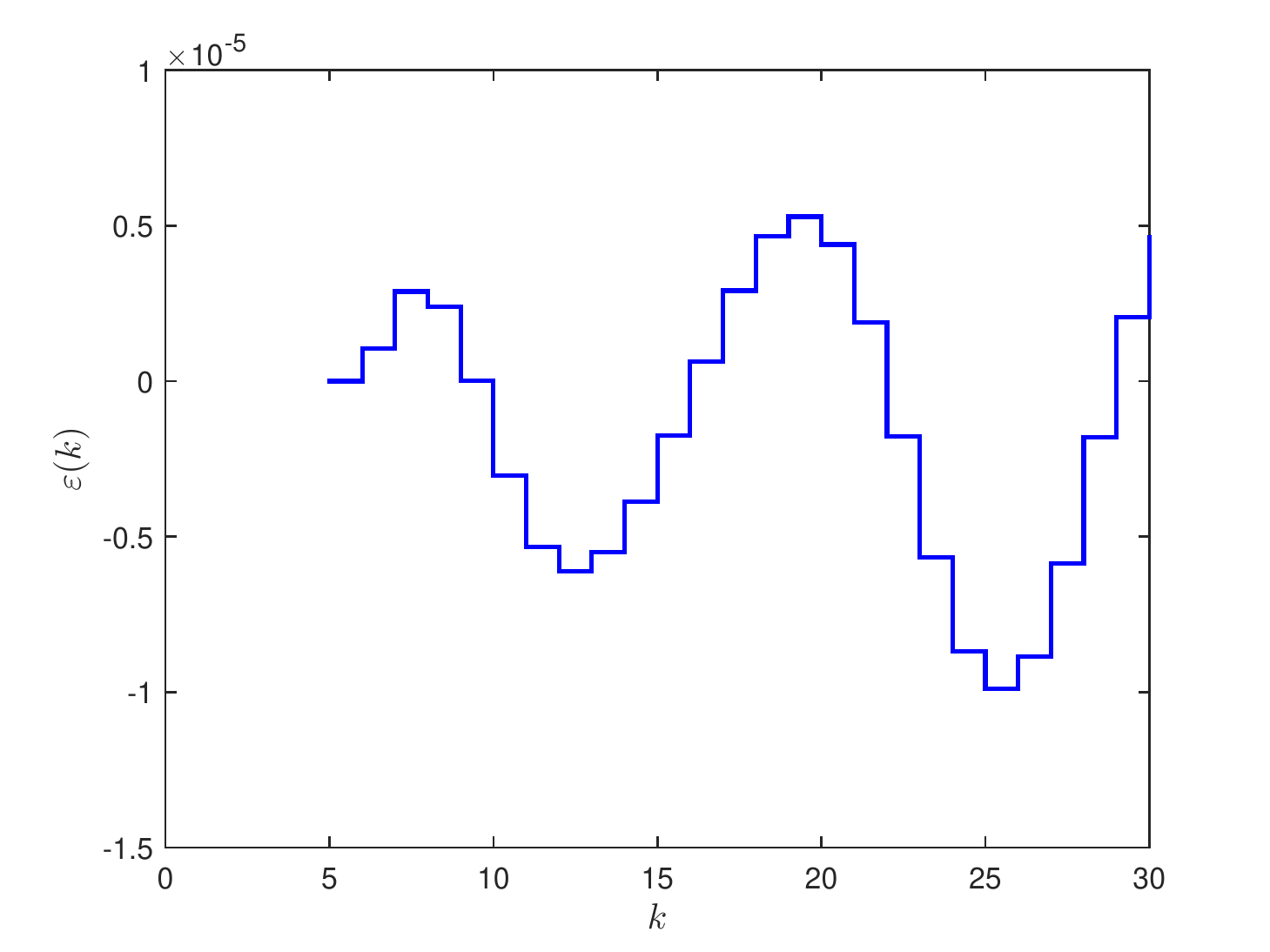}
  \caption{The approximate error $\varepsilon(k)$.}\label{Fig6}
\end{figure}

Consider a system
\begin{equation}\label{Eq41}
{}_a^{}\nabla_k^{\alpha}y(k)=-2 y(k)+u(k),
\end{equation}
where the system input $ u(k)=5\sin(0.2\pi k)$ is shown as Fig. \ref{Fig4}. With the similar method in Example 1, the analytical output and the simulated output are displayed in Fig. \ref{Fig5}. The error between the two outputs is represented in Fig. \ref{Fig6}. The relative error is inferior to $10^{-5}$, which is quite acceptable.


\vspace{20pt}
\noindent\textbf{Example 3: For the nonlinear system}

\begin{figure}[htbp]
  \centering
  \includegraphics[width=0.5\textwidth]{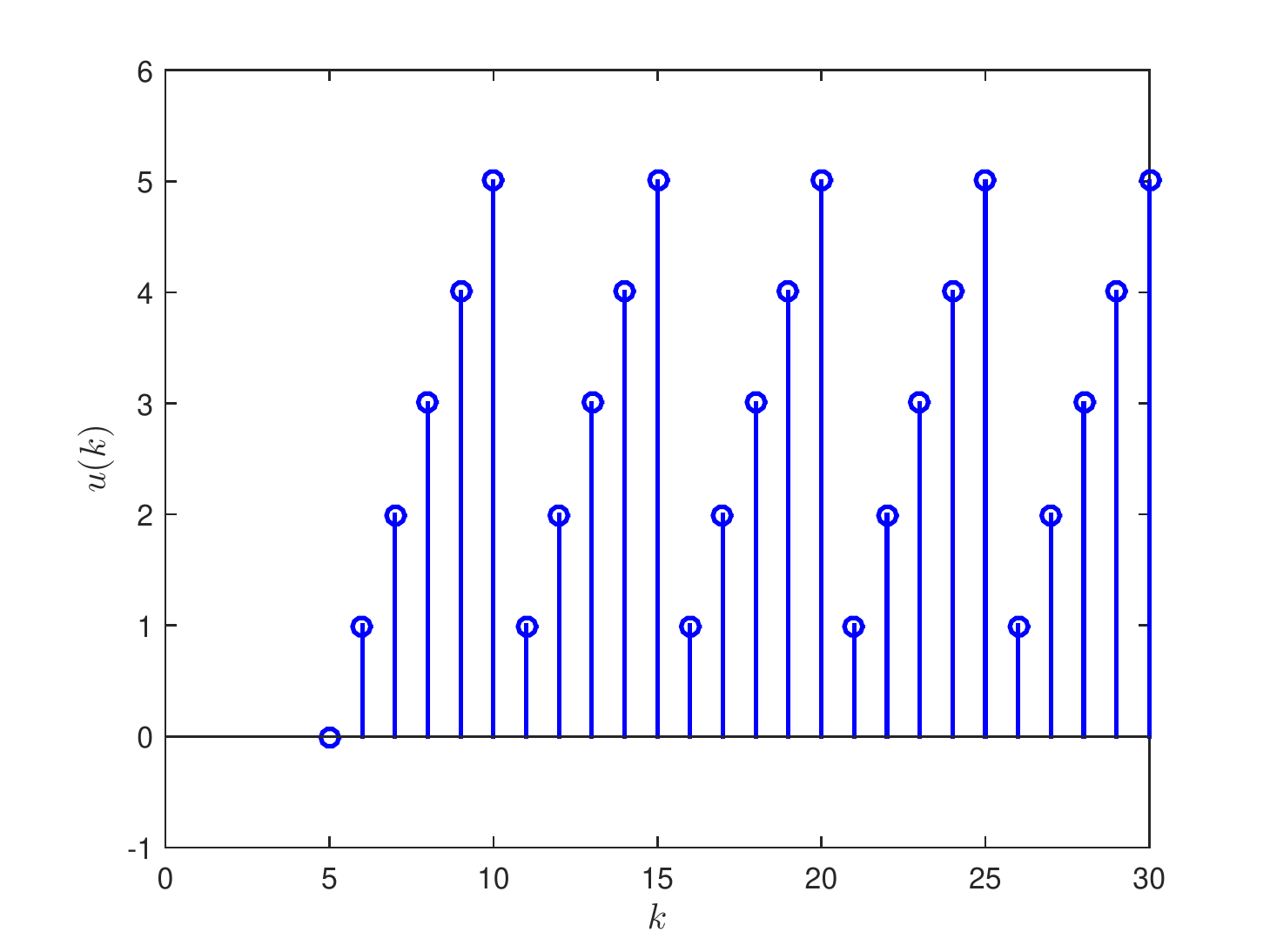}
  \caption{The system input  $u(k)$.}\label{Fig7}
\end{figure}

\begin{figure}[htbp]
  \centering
  \includegraphics[width=0.5\textwidth]{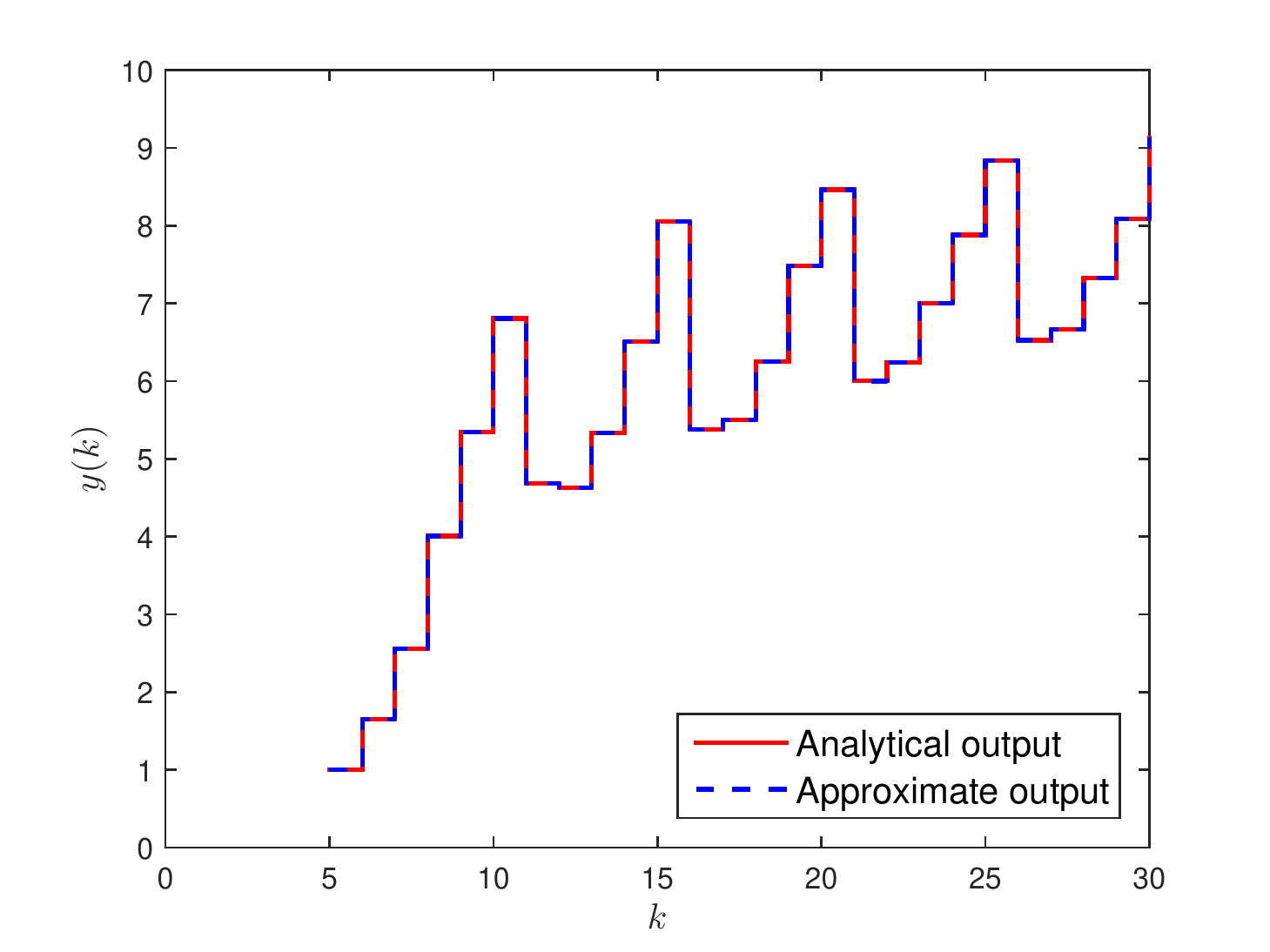}
  \caption{The system output $y(k)$.}\label{Fig8}
\end{figure}

\begin{figure}[htbp]
  \centering
  \includegraphics[width=0.5\textwidth]{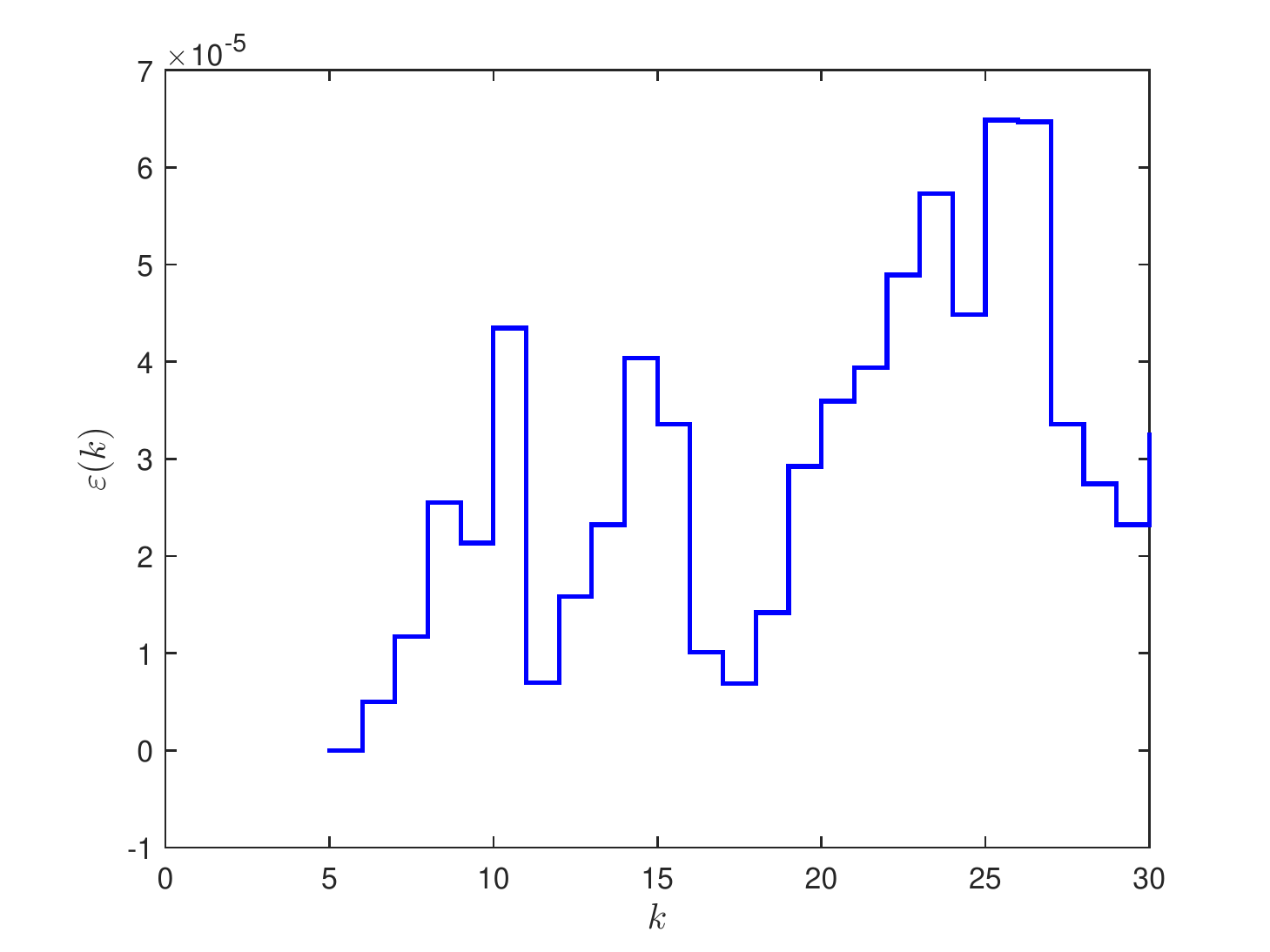}
  \caption{The approximate error $\varepsilon(k)$.}\label{Fig9}
\end{figure}

Consider a nonlinear fractional dynamic system
\begin{eqnarray}\label{Eq42}
{}_a^{}\nabla_k^{\alpha}y(k)=-0.3 y(k)+0.5{\cos^2(y(k-1))}+u(k),
\end{eqnarray}
where the input $u(k)$ is a sawtooth wave signal shown as Fig. \ref{Fig7}. The period is 5 and the amplitude is 5. In general, it is difficult to get the analytical output of a nonlinear system. To verify the effectiveness of the established approximate method, the nonlinear term is selected regarding to $y(k- 1)$ not $y(k)$. The outputs $y(k)$ solved by the definition and the approximate method are displayed in Fig. \ref{Fig8}. The approximate error is provided in Fig. \ref{Fig9}.


It is shown that the approximate error is quite small, which illustrates that the developed scheme is very effective and efficient and the approximate results can reflect the dynamic performance of the original system to a large degree. Without loss of generality, the discussed method can be adopted in the simulation of the general nonlinear case and other different nabla fractional order systems.

\section{Conclusions}\label{Section 5}
In this paper, the numerical approximation has been investigated for nabla fractional dynamic systems with nonzero initial instant and nonzero initial state. According to the frequency distributed model theory, the exact infinite dimensional model of fractional sum operator is recalled and then the approximation problem is transformed into a nonlinear identification problem for the first time. By applying the vector fitting approach in frequency domain, an innovative and effective numerical approximation is developed. Finally, three numerical examples have verified the general applicability and flexibility of the proposed results. It is believed that the developed method will play an essential role in the applications of discrete fractional calculus.

%
% ...

\bibliographystyle{asmems4}

\small
\bibliography{database}
\end{document}